\documentclass[longbibliography,showpacs,superscriptaddress,amsmath,amssymb,10pt,pra,aps,floatfix,twocolumn]{revtex4-1}
\pdfoutput=1 

\usepackage{graphicx}
\usepackage{epsfig}
\usepackage{color}
\usepackage{hyperref}
\usepackage{verbatim}
\usepackage{xfrac}
\usepackage{braket}
\usepackage{multirow}

\begin{document}
%
%
%
%
	\title{Symmetry breaking and phase transitions in Bose-Einstein condensates with spin-orbital-angular-momentum coupling}
%
%
%
%

	\author{Y.~Duan}
	\affiliation{Physikalisch--Technische Bundesanstalt, D--38116 Braunschweig, Germany}
	\affiliation{Institut f\"ur Mathematische Physik, Technische Universit\"at Braunschweig, D--38106 Braunschweig, Germany}

	\author{Y.~M.~Bidasyuk}
	\affiliation{Physikalisch--Technische Bundesanstalt, D--38116 Braunschweig, Germany}
	
	\author{A.~Surzhykov}
	\affiliation{Physikalisch--Technische Bundesanstalt, D--38116 Braunschweig, Germany}
	\affiliation{Institut f\"ur Mathematische Physik, Technische Universit\"at Braunschweig, D--38106 Braunschweig, Germany}
	\affiliation{Laboratory for Emerging Nanometrology Braunschweig, D--38106 Braunschweig, Germany}
	
	\date{\today \\[0.3cm]}

%
%
\begin{abstract}
    Theoretical study is presented for a spinor Bose-Einstein condensate, whose two components are coupled by copropagating Raman beams with different orbital angular momenta.
    The investigation is focused on the behavior of the ground state of this condensate, depending on the atom-light coupling strength. By analyzing the ground state, we have identified a number of quantum phases, which reflect the symmetries of the effective Hamiltonian and are characterized by the specific structure of the wave function. 
    In addition to the well-known stripe, polarized and zero-momentum phases, our results show that 
    the system can support phases, whose wave function contains a complex vortex molecule. Such molecule plays an important role in the continuous phase transitions of the system. 
    The predicted behavior of vortex-molecule phases can
    be examined in cold-atom experiments using currently existing techniques.
\end{abstract}

\maketitle

\section{Introduction}\label{introduction}

The first experimental realizations of atomic Bose-Einstein condensates (BEC) in 1995 opened a new era in the study of macroscopic quantum systems \cite{Anderson198,PhysRevLett.75.3969}.
In the original experiments, the condensation of atoms in a single hyperfine sublevel was achieved using a magnetic trap. Further developments of optical-dipole trapping techniques enabled the simultaneous confinement of atoms in several hyperfine substates and thus opened a possibility to manipulate also the spin degree of freedom of the condensate \cite{PhysRevLett.80.2027,KAWAGUCHI2012253}.
Coherent coupling between the substates of such spinor condensate is commonly achieved with two-photon lambda-type Raman transitions. 
If the Raman transition is accompanied by a change in the center-of-mass motion of an atom, the spin and motional degrees of freedom become coupled.
This artificial spin-orbit coupling (SOC) has drawn considerable attention from both theory and experiment \cite{Lin2011,PhysRevLett.109.095301,RevModPhys.83.1523,Zhai2015,Goldman_2014,ZHANG201975}. 
In particular, previous studies demonstrated the utility of atomic BEC as a highly controllable platform for exploring effects of SOC in various fields of modern physics \cite{ZHANG201975}.
As the most prominent example, spin-orbit-coupled condensate has been proposed to simulate such exotic phenomena in condensed matter physics as topological insulators and quantum spin Hall effect \cite{PhysRevA.93.053605,beeler2013spin}.
Moreover, BEC with SOC can be seen as a testbed for studying quantum phase transitions at zero temperature, as it supports a number of quantum phases with distinct symmetries and topological properties \cite{PhysRevLett.108.225301,PhysRevLett.105.160403,PhysRevA.86.063621,PhysRevLett.110.235302,Li2017}.

So far, different experimental setups have been proposed to produce spin-orbit-coupled condensates \cite{ZHANG201975}. In the first realization, for example, counterpropagating Raman lasers were used to enable a transfer of linear momentum to atoms during the ``absorption-and-stimulated-emission" process \cite{Lin2011}. In addition to this spin-linear-momentum (SLM) coupling, another type of SOC can be induced by two copropagating beams with different orbital angular momenta \cite{PhysRevA.91.033630}. This setup suppresses the transfer of linear momentum, and couples instead the spin and orbital angular momentum (OAM) of the condensate. BECs with the spin-orbital-angular-momentum (SOAM) coupling have been studied theoretically \cite{PhysRevA.93.013629,PhysRevA.94.033627,PhysRevA.91.063627,PhysRevA.92.033615,PhysRevA.91.053630,PhysRevA.91.033630,PhysRevResearch.2.033152,PhysRevA.102.013316}, and realized recently in the experiments with $^{87}\mathrm{Rb}$ atoms \cite{PhysRevLett.121.113204,PhysRevLett.122.110402}. 
In these studies, special attention has been paid to their rich ground-state phase diagram. In particular, it has been shown that several quantum phases in SOAM coupled condensates are very similar to those observed under SLM coupling \cite{PhysRevA.91.063627,PhysRevLett.122.110402}.
However, a few other phases, which have no obvious counterparts in the SLM scenario, were also predicted \cite{PhysRevA.92.033615, PhysRevA.91.053630,PhysRevResearch.2.033152,PhysRevA.102.013316}. These new phases are characterized by the existence of quantum vortices or other topological defects.
A number of theoretical studies have been performed to search for such vortex phases and to analyze the complex topological structure of their wave functions \cite{PhysRevA.92.033615, PhysRevA.91.053630,PhysRevResearch.2.033152,PhysRevA.102.013316}.  
Not so much attention, however, has been paid to the questions of how the vortex states change through the phase transitions and how these changes are related to the \emph{symmetries} of the system.
As we show below, answers to these questions are especially important for understanding the microscopic mechanism driving the phase transitions in SOAM-coupled atomic condensates.  

In the present work, we aim to analyze tight interrelations between the quantum phase transitions and symmetries in spinor Bose-Einstein condensates with SOAM coupling.
To this end, we first define in Sec.~\ref{sec:theoretical} our model system. We consider a quasi-two-dimensional condensate whose two internal levels are resonantly coupled by copropagating Laguerre-Gaussian laser modes.
The Gross-Pitaevskii Hamiltonian of this system as well as the properties of its eigenstates are discussed in Secs.~\ref{model} and \ref{sub:symmetry}.
The numerical procedure used to reliably compute the ground-state solutions of the Hamiltonian is briefly explained in Sec.~\ref{sub:numerical}.
This procedure is employed later in Sec.~\ref{sec:results} for a realistic scenario of  $^{87}\mathrm{Rb}$ condensate in a disk-shaped harmonic trapping potential. In particular, the ground-state wave function is calculated for various values of the Raman coupling strength. Based on the results of these calculations, we identify in total five quantum phases of the spinor condensate. We describe these phases and analyze the relation between their vortex structures and symmetry properties.
Moreover, we argue that the quantum vortices in the condensate can strongly affect the mechanism of phase transitions. Particularly, we find that the counterintuitive continuous phase transitions, predicted for the SOAM coupled condensate, can be explained by the formation of ``vortex molecule" in the ground-state wave function.
The summary and outlook are given in Sec.~\ref{summary}.

\section{Theoretical framework}\label{sec:theoretical}
\subsection{Model}\label{model}
\begin{figure}
    \centering
    \includegraphics[trim=0 0 0 0,width=\linewidth,clip]{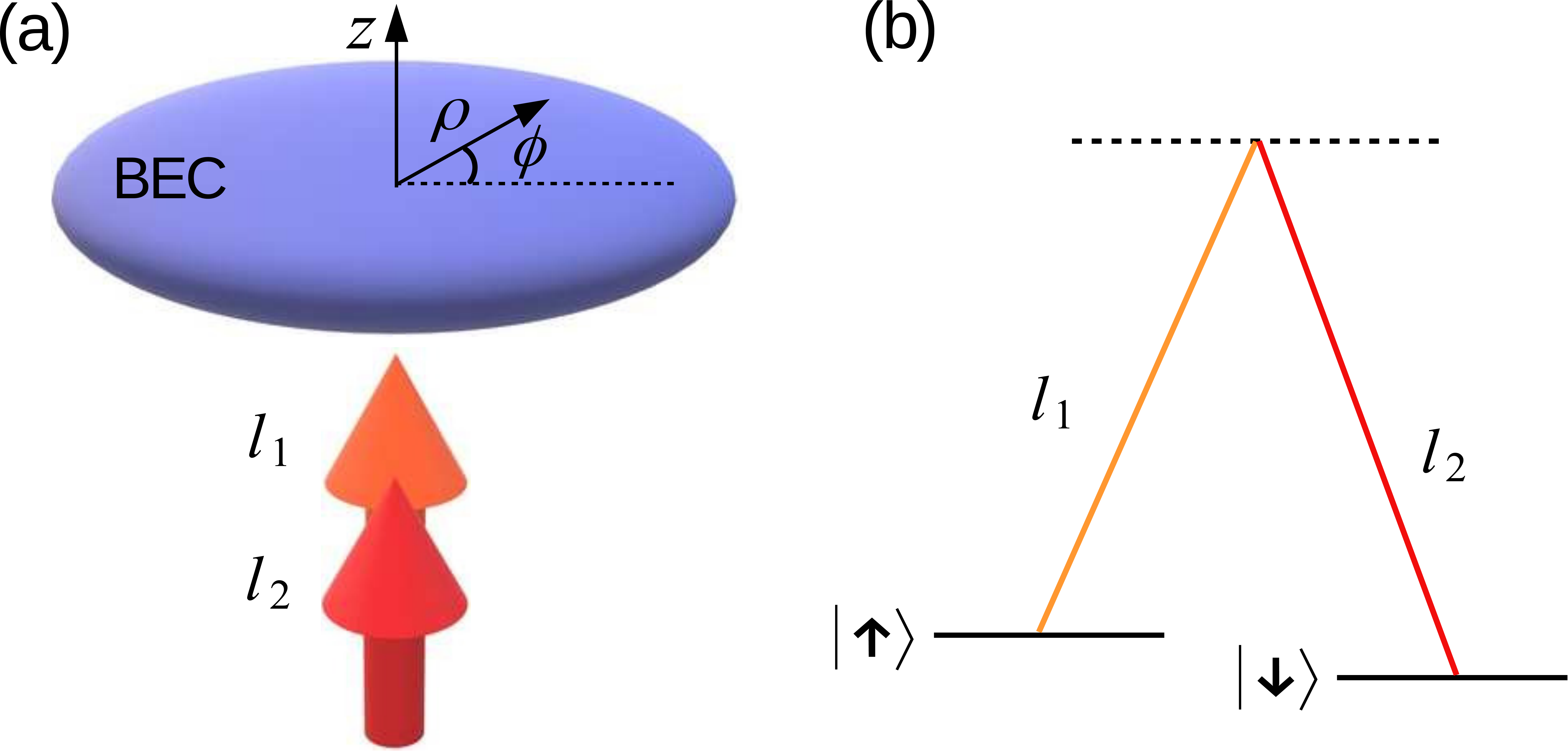}
	\caption{(a) Schematic representation of a disk-shaped BEC interacting with copropagating LG laser beams with OAM $l_1$ and $l_2$. (b) A sketch of the level diagram illustrating the angular momentum transfer in the two-photon Raman transition between atomic states $\ket{\uparrow}$ and $\ket{\downarrow}$
	}
    \label{fig:geometry}
\end{figure}

We consider a weakly interacting Bose-Einstein condensate in a harmonic trap. The condensate atoms are assumed to occupy two internal atomic states, coupled by laser beams in a lambda-type Raman regime [see Fig.~\ref{fig:geometry}(b)]. For simplicity we denote these two states as $\ket{\uparrow}$ and $\ket{\downarrow}$. The condensate is then described by a coherent superposition of two macroscopic wave functions as 
\begin{equation}\label{spinorwavefunction}
    \Psi(\mathbf{r})=\psi_\uparrow(\mathbf{r})\chi_\uparrow+\psi_\downarrow(\mathbf{r})\chi_{\downarrow}.
\end{equation}
For convenience of representation, we consider $\chi_\uparrow=\left(\begin{smallmatrix}1\\0\end{smallmatrix}\right)$ and $\chi_\downarrow=\left(\begin{smallmatrix}0\\1\end{smallmatrix}\right)$ as the basis of a two-dimensional complex Hilbert space. In such representation, $\Psi(\mathbf{r})$ can be written as a two-component vector
\begin{equation} \label{eq:generalwavefunction}
    \Psi(\mathbf{r})=\left[\begin{matrix}
	\psi_\uparrow(\mathbf{r}) \\  \psi_\downarrow(\mathbf{r})
	\end{matrix}\right].
\end{equation}
This pseudo-spin-$\sfrac{1}{2}$ wave function $\Psi$ obeys the Gross-Pitaevskii equation (GPE): 
\begin{equation}
	i\hbar\frac{\partial}{\partial t}\Psi=\hat{H}\Psi\equiv(\hat{H}_\mathrm{s}+\hat{H}_\mathrm{int})\Psi\,,
	\label{gpe}
\end{equation} 
where the Hamiltonian $\hat{H}$ is a $2\times2$ matrix consisting of single-particle and nonlinear interaction terms \cite{PhysRevA.91.053630}. The first one
\begin{equation}\label{eq:ham}
	\hat{H}_\mathrm{s}=-\frac{\hbar^2\nabla^2}{2M}\mathbb{I}_2+V_\mathrm{trap}\mathbb{I}_2+\hat{V}_\mathrm{c}
\end{equation}
is a sum of kinetic energy operator, trapping potential and the Raman coupling $\hat{V}_\mathrm{c}$ between two internal states. The interaction Hamiltonian $\hat{H}_\mathrm{int}$ in Eq.~(\ref{gpe}) reads as
\begin{equation}\label{eq:nonlinear}
	\hat{H}_\mathrm{int}=\left(\begin{matrix}
	g_{\uparrow\uparrow}|\psi_\uparrow|^2+g_{\uparrow\downarrow}|\psi_\downarrow|^2 & 0 \\  0 & g_{\downarrow\downarrow}|\psi_\downarrow|^2+g_{\uparrow\downarrow}|\psi_\uparrow|^2
	\end{matrix}\right)\,,
\end{equation}  
where $g_{\uparrow\uparrow}$, $g_{\downarrow\downarrow}$ and $g_{\uparrow\downarrow}$ are nonlinear interaction parameters characterizing collisions between particles in corresponding internal states.

To further specify the single-particle Hamiltonian (\ref{eq:ham}), we need to define the geometry of our system. In particular, we consider the condensate to be confined in a disk-shaped harmonic trap as illustrated in Fig.\ref{fig:geometry}(a). In cylindrical coordinates $\mathbf{r}=(\rho,\phi,z)$, the trap potential is described by:
\begin{equation}
    V_\mathrm{trap}(\mathbf{r})=\frac{M\omega_\rho^2\rho^2}{2}+\frac{M\omega_z^2z^2}{2},
\end{equation}
where $\omega_z$ and $\omega_\rho$ are the trapping frequencies in $z$ and radial directions, respectively. We assume $\omega_z \gg \omega_\rho$, which allows to consider $z$ dimension as frozen and treat the system as two-dimensional
\cite{PhysRevA.65.043617,PhysRevLett.122.110402}. 

As it is schematically shown in Fig.\ref{fig:geometry}, a resonant Raman coupling  between two atomic states is induced by two Laguerre-Gaussian (LG) lasers with different OAM $l_1$ and $l_2$, which copropagate along $z$ axis. 
The coupling term $\hat{V}_\mathrm{c}$ in Eq.~(\ref{eq:ham}) is then defined by the effective two-photon Rabi frequency \cite{PhysRevA.91.063627}. In the paraxial regime, it reads as
\begin{equation}\label{eq:coupling}
	\hat{V}_\mathrm{c}(\mathbf{r}) = \tilde\Omega(\rho)\left(\begin{matrix}
	0 & e^{2il\phi} \\ e^{-2il\phi} & 0
	\end{matrix}\right) \equiv \tilde\Omega(\rho)\hat{\sigma}_{2l}\,,
\end{equation}
where $2l=l_2-l_1$ and the unitary operator $\hat{\sigma}_{2l}$ can be represented in terms of the Pauli matrices $\hat{\sigma}_{x}$ and $\hat{\sigma}_{z}$: \begin{equation}
    \hat{\sigma}_{2l}=e^{il\phi\hat{\sigma}_z}\hat{\sigma}_xe^{-il\phi\hat{\sigma}_z}.
\end{equation}
Moreover, in Eq.~(\ref{eq:coupling}), $\tilde\Omega(\rho)$ is the radial function, which reflects the intensity distribution of the LG modes and is given by
\begin{equation}\label{eq:radial}
  \tilde\Omega(\rho)=\Omega\, \left(\frac{\rho}{w}\right)^{2l}e^{-\frac{2\rho^2}{w^2}}, 
\end{equation}
where $w$ is the beam waist and $\Omega$ is the amplitude parameter proportional to the beam intensity, which we call the \emph{Raman coupling strength}. 

To complete the introduction of the theoretical model, we should mention two integrals of motion of the GPE (\ref{gpe}). These are the total particle number $N$, which we choose for the normalization of the wave functions
\begin{equation}
    N = \int {\rm d}\mathbf{r} \left(|\psi_\uparrow|^2+|\psi_\downarrow|^2\right),
\end{equation}
and the energy per particle
\begin{eqnarray}\label{eq:energy}
    \epsilon = \frac{1}{N}\int {\rm d}\mathbf{r} \left[\left(\psi^*_\uparrow,\psi^*_\downarrow\right)\hat{H}_\mathrm{s}\left(\psi_\uparrow,\psi_\downarrow\right)^{T} \right.\nonumber\\
    \left.+\frac{g_{\uparrow\uparrow}}{2}|\psi_\uparrow|^4+\frac{g_{\downarrow\downarrow}}{2}|\psi_\downarrow|^4+g_{\uparrow\downarrow}|\psi_\uparrow|^2|\psi_\downarrow|^2\right].
\end{eqnarray}
The latter expression allows to define the ground state of the system as a state corresponding to the lowest possible value of $\epsilon$.

\subsection{Symmetry properties of the system}\label{sub:symmetry}

It is convenient to begin the analysis of the GPE (\ref{gpe}) by addressing the non-interacting case $\hat H_{\mathrm{int}}=0$.
The GPE in this case becomes a single-particle Schr\"odinger equation.
In order to characterize its solutions we need to analyze relevant symmetries defined by transformation operators commuting with the Hamiltonian $\hat H_s$. For a system with SOAM coupling the relevant symmetry transformations include rotations and time reversal.

We first analyze the rotational symmetry with respect to $z$ axis. 
Due to the $\phi$-dependence of Raman coupling $\hat{V}_c$, the single-particle Hamiltonian (\ref{eq:ham}) is not invariant under rotation of spatial coordinates, which is generated by the operator
\begin{equation}
\hat{R}(\phi_0) = e^{-i\hat{L}_z\phi_0}, \,\,\,\text{with}\,\,\,\, \hat{L}_z = -i\frac{\partial}{\partial\phi}\mathbb{I}_2.
\end{equation}
This means, in particular, that orbital angular momentum projection $\langle \hat{L}_z \rangle=\frac{1}{N}\bra{\Psi}\hat{L}_z\ket{\Psi}$ is not conserved in the system. 
On the other hand, the operator that combines spatial and spin rotation, 
\begin{equation}\label{eq:rot-op}
\hat{R}'(\phi_0) = e^{-i\left(\hat{L}_z-l\hat{\sigma}_z\right)\phi_0},
\end{equation}
commutes with the Hamiltonian (\ref{eq:ham}) and represents therefore a symmetry of the system. This symmetry transformation implements the spatial rotation by the angle $\phi_0$ with a simultaneous spin rotation by the angle $-l\phi_0$. We expect therefore the conservation of the ``total'' angular momentum (TAM) projection defined by the operator
\begin{equation}\label{eq:tam}
    \hat{J}_z = \hat{L}_z - l \hat{\sigma}_z.
\end{equation}
One may notice that a conceptual similarity of our system to a standard spin-orbit coupling in atomic and nuclear physics.
In contrast to atoms and nuclei, however, the eigenvalues of $\hat J_z$ for SOAM-coupled condensates span the entire range of integer values $-\infty < j < \infty$. 

The eigenstates of the Hamiltonian (\ref{eq:ham}) with well-defined TAM projection can be written in the general form:
\begin{equation}\label{eq:jeigf}
    \Psi_j(\mathbf{r}) = \left[\begin{matrix}
         f_j(\rho) e^{i(j+l)\phi}  \\
         g_j(\rho) e^{i(j-l)\phi}
    \end{matrix} \right],
\end{equation}
where the radial functions $f_j(\rho)$ and $g_j(\rho)$ can be considered real-valued without a loss of generality \cite{PhysRevA.92.033615}.
In the discussion below we will refer to the symmetry associated with the operator (\ref{eq:rot-op}) as \emph{rotational symmetry} ($\mathrm{R}$-symmetry) and to the states (\ref{eq:jeigf}) as \emph{rotationally-symmetric} states of the system. 
To identify these states, we can evaluate the symmetry indicators, given by the expectation value
\begin{equation}
    \langle \hat{J}_z\rangle=\frac{1}{N}\bra{\Psi}\hat{J}_z\ket{\Psi}   
\end{equation}
and a standard deviation 
\begin{equation}
\Delta J_z \equiv \sqrt{\langle\hat{J}_z^2\rangle - \langle\hat{J}_z\rangle^2}.
\end{equation}
By inserting $\Psi=\Psi_j$ into above equations, we obtain $\langle \hat{J}_z\rangle=j$ and $\Delta J=0$.
For a general non-R-symmetric state, we can expect $\Delta J>0$. This measure will be used as a quantitative indicator of the rotational symmetry in our numerical calculations below.

As already mentioned above, another relevant symmetry transformation is the time reversal.
The Hamiltonian $\hat{H}_s$ obviously commutes with the time-reversal operator $\hat{T}=\hat{\sigma}_x\hat{K}$, with $\hat{K}$ being the complex conjugation. 
Eigenfunctions of this operator, which correspond to real eigenvalues $\pm1$, have the following structure:
\begin{equation}\label{eq:teigf}
    \Phi_{\pm}(\mathbf{r}) = \left[\begin{matrix}
         \psi(\mathbf{r}) \\
         \pm\psi^*(\mathbf{r}) 
    \end{matrix} \right].
\end{equation}
Here $\psi(\mathbf{r})$ is an arbitrary function which has no immediate relation to the functions $f(\rho)$ and $g(\rho)$ in Eq.~(\ref{eq:jeigf}).
We will refer to the states  (\ref{eq:teigf}) as \emph{time-reversal-symmetric}.
Again, we would like to have a convenient measure which quantifies the time-reversal symmetry of an arbitrary wave function (\ref{eq:generalwavefunction}).
In contrast to the rotational symmetry, however, there is no physical observable associated with the operator $\hat T$. One suitable quantity, commonly applied for spinor condensates and related to $\mathrm{T}$-symmetry, is the spin polarization 
\begin{equation}
    \langle\hat{\sigma}_z\rangle=\frac{1}{N}\int (|\psi_\uparrow(\mathbf{r})|^2-|\psi_\downarrow(\mathbf{r})|^2)\mathrm{d}\mathbf{r},
\end{equation}
which represents the relative population imbalance between two components of the wave function \cite{PhysRevLett.122.110402,PhysRevA.91.053630}. This quantity, however, is not a universal indicator of the time-reversal symmetry. As one can see from Eq.~(\ref{eq:teigf}), time-reversal-symmetric states are always unpolarized, $\langle\hat{\sigma}_z\rangle=0$. The opposite however is not true, since the unpolarized state can be also non-T-symmetric. We will encounter such non-symmetric unpolarized states in the numerical results of the next section.

Above we have analyzed each of the two symmetries of the system separately. 
We now discuss the consequences of both of them being applied simultaneously.
We first remind that
the angular momentum operator transforms under time reversal as $\hat T \hat J_z \hat T^{-1} = - \hat J_z$. 
Consequently, the Hamiltonian eigenstates with opposite values of the TAM projection are related as 
\begin{equation}\label{eq:symrel}
    \hat T \Psi_j(\mathbf{r}) = \pm \Psi_{-j}(\mathbf{r}).  
\end{equation}
The time-reversal symmetry of the Hamiltonian also implies that both $\Psi_j(\mathbf{r})$ and $\hat T \Psi_j(\mathbf{r})$ are eigenstates with the same energy. 
Since the states $\Psi_j(\mathbf{r})$ and $\Psi_{-j}(\mathbf{r})$ are orthogonal for $j \neq 0$, then corresponding energy levels are doubly degenerate.
The state with $j=0$ is nondegenerate and possesses both symmetries. This additionally restricts the radial parts of the corresponding wave function to $g_0(\rho) = \pm f_0(\rho)$.

The above considerations are valid for all eigenstates of the Hamiltonian $\hat H_s$. 
However, our primary goal here is to characterize the properties of the ground state.
According to the previous studies \cite{PhysRevA.94.033627,PhysRevA.91.053630,PhysRevA.92.033615}, the ground state of a SOAM-coupled system may have either zero or non-zero angular momentum projection $j_0$ within the range $-l \leq j_0 \leq l$.
If the ground state corresponds to zero angular momentum $j_0=0$ then it is also $\mathrm{T}$-symmetric as already discussed above.
For the cases when $j_0 \neq 0$, there are two degenerate rotationally-symmetric ground states $\Psi_{j_0}$ and $\Psi_{-j_0}$. Their superposition  allows to define equivalent time-reversal-symmetric ground states as 
\begin{equation}\label{eq:superosition}
\Phi_{j_0 \pm}=\frac{1}{\sqrt2}\left(\Psi_{j_0} \pm \Psi_{-j_0}\right).    
\end{equation}
These two ground states are not rotationally-symmetric and characterized by $\langle J_z \rangle = 0$ and $\Delta J_z = j_0$.
Analogous discussion for the case of SLM coupling can be found in Ref.~\cite{PhysRevLett.105.160403}.

So far we have discussed symmetry properties of the non-interacting Hamiltonian.
Most of them are expected to be valid also in the presence of collisional interactions manifested by the nonlinear term in the Hamiltonian (\ref{eq:nonlinear}).
However, one crucial difference of the interacting system is a breakdown of the superposition principle for wave functions.
This superposition principle is required for R-symmetric states $\Psi_{\pm j_0}$ and T-symmetric states $\Phi_{j_0\pm}$ to have the same energy. In the interacting system their energies will be different.
Therefore, if the ground state is characterized by $j_0 \neq 0$, then
either rotational or time-reversal symmetry will be spontaneously broken. These symmetry breakings lead to possible existence of two different quantum phases: a stripe phase with broken R-symmetry, and a polarized phase with broken T-symmetry \cite{PhysRevLett.108.225301,PhysRevA.91.053630}.
If the ground state is characterized by $j_0=0$ then both symmetries can be preserved even in the presence of  interactions. Such type of ground states of the interacting systems is commonly termed as zero-momentum phase.
These and other possible types of ground states of the interacting system will be discussed in more detail in the next section.

\subsection{Numerical procedure}\label{sub:numerical}
Symmetries of the single-particle Hamiltonian provide a lot of insight into the properties of the ground state. Nevertheless, solutions of the full GPE (\ref{gpe}) can never be obtained analytically. 
For numerical computations of the ground-state solution 
we adopt a well-known imaginary-time evolution method 
\cite{LEHTOVAARA2007148,doi:10.1137/S1064827503422956}.
Due to a complex energy landscape of a SOAM-coupled system, finding a true ground state becomes a numerically challenging task.
To distinguish the ground state from low-lying metastable excited states, we repeat the procedure of imaginary-time evolution using different initial trial states.
We construct trial states with different populations and OAM projections between $-2l$ an $2l$ in each component, as well as states with random distributions.
The ground state is then determined as the final state with the lowest energy after the converged numerical procedure.

\section{Results and discussions}\label{sec:results}

For the rest of the paper we focus on 
a more realistic system of weakly interacting atoms described by the full GPE (\ref{gpe}).
We consider a total number of $N=5\times10^3$ particles confined in a disk-shaped harmonic trap with trapping frequencies $\omega_\rho=1.5\times 2\pi~\mathrm{Hz}$ and $\omega_z=24\times 2\pi~\mathrm{Hz}$. For this choice of frequencies, $\omega_z\gg\omega_\rho$, the use of two-dimensional approximation 
is justified. 
The nonlinear interaction parameters are chosen as
\[
g_{\uparrow\uparrow} = g_{\downarrow\downarrow} = g, \quad
g_{\uparrow\downarrow} = 0.9 g.
\]
where $g=a\sqrt{8\pi\hbar^2\omega_z/M}$ is related to the s-wave scattering length $a$ and the atom mass $M$ of $^{87}$Rb.
This choice of nonlinear interaction parameters is reasonably close to the realistic parameters of $^{87}$Rb condensate but at the same time allows to observe many possible types of the ground state  \cite{PhysRevResearch.2.033152}. 

The Raman coupling (\ref{eq:coupling}) is characterized by the angular momentum $l=1$ and the beam waist $w=176~\mathrm{\mu m}$, which is considerably larger than the estimated Thomas-Fermi radius of the condensate $R_{\mathrm{TF}} \approx 26~\mathrm{\mu m}$. 

Our results below are presented in dimensionless units, by adopting the oscillator energy $\hbar\omega_\rho$ and length $\sqrt{\hbar/M\omega_\rho}$ of the harmonic trap as the energy and length scales, respectively.

\subsection{Phase diagram}\label{sub:phasediagram}

\begin{figure}[tbp]
	\includegraphics[width=\linewidth]{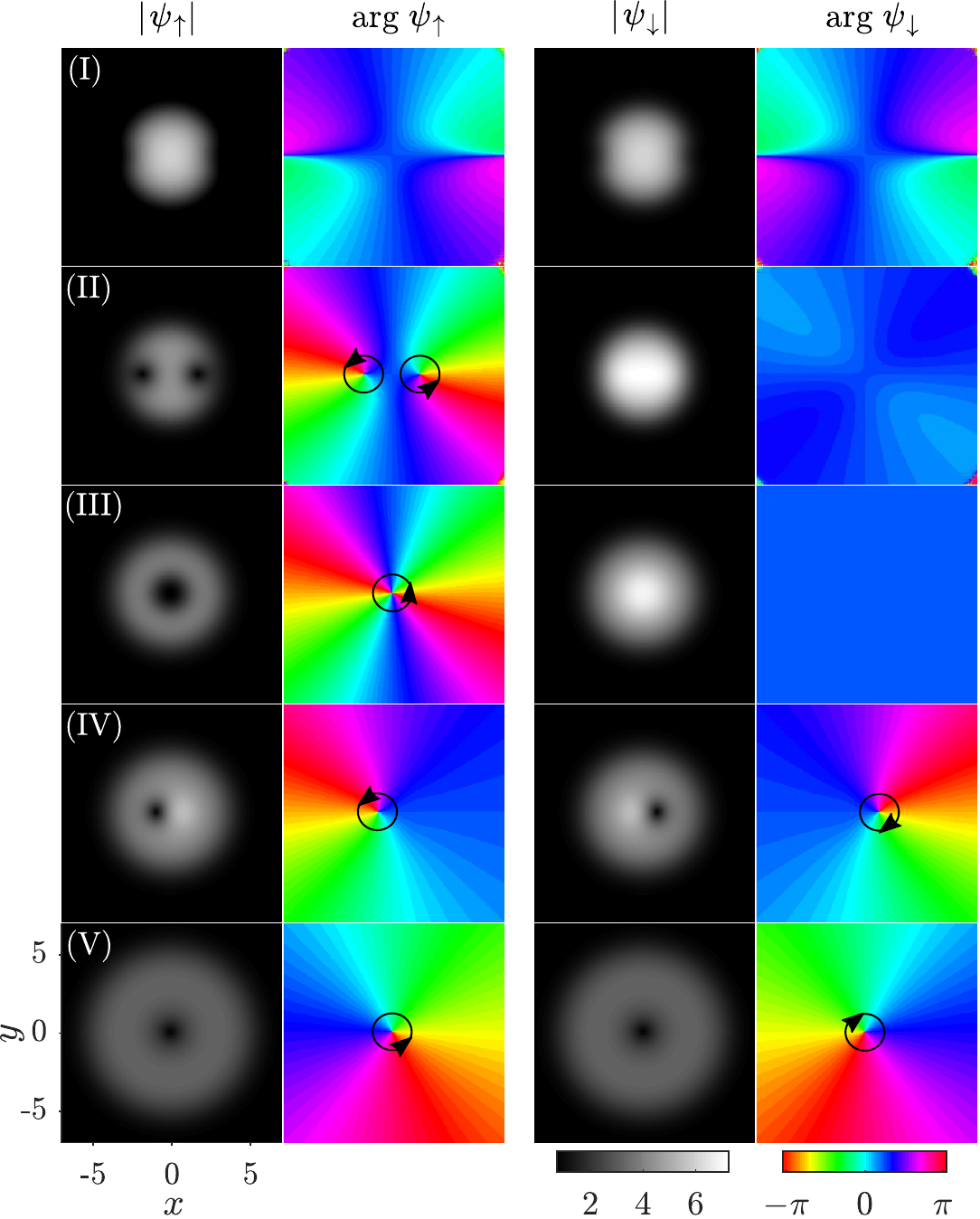}
	\caption{Amplitudes and phases of the ground state wave functions for up (two left columns) and down (two right columns) components at Raman coupling strengths $\Omega=36$ (I), $\Omega=45$ (II), $\Omega=120$ (III), $\Omega=150$ (IV) and $\Omega=220$ (V). Vortices in each component are marked with circles on the phase plots. Arrows indicate directions of the phase winding.}
	\label{fig:5phases}
\end{figure}

We aim here to define and characterize possible types of ground states in the system.
Due to a large number of parameters in the Hamiltonian, it is unfeasible to explore the entire parameter space. 
We therefore concentrate on a single, but most relevant control parameter, the Raman coupling strength $\Omega$.
Fig.~\ref{fig:5phases} shows five examples of ground-state wave functions calculated for different values of $\Omega$.
One may see qualitative changes in the ground state depending on the coupling strength.
When the coupling is weak, we can observe a deformation of the cloud from the rotationally-symmetric shape defined by the trapping potential [see Fig.~\ref{fig:5phases}(I)]. 
Stronger coupling leads to the formation of quantum vortices in the condensate, which can be identified by hollow regions in the amplitude and singularities in the phase of the wave function. 
By calculating the ground-state solutions in a wide range of $\Omega>0$
we are able to identify four different spatial arrangements of vortices in the condensate [see Fig.~\ref{fig:5phases}(II--V)]. 

\begin{figure}[tbp]
	\includegraphics[width=\linewidth]{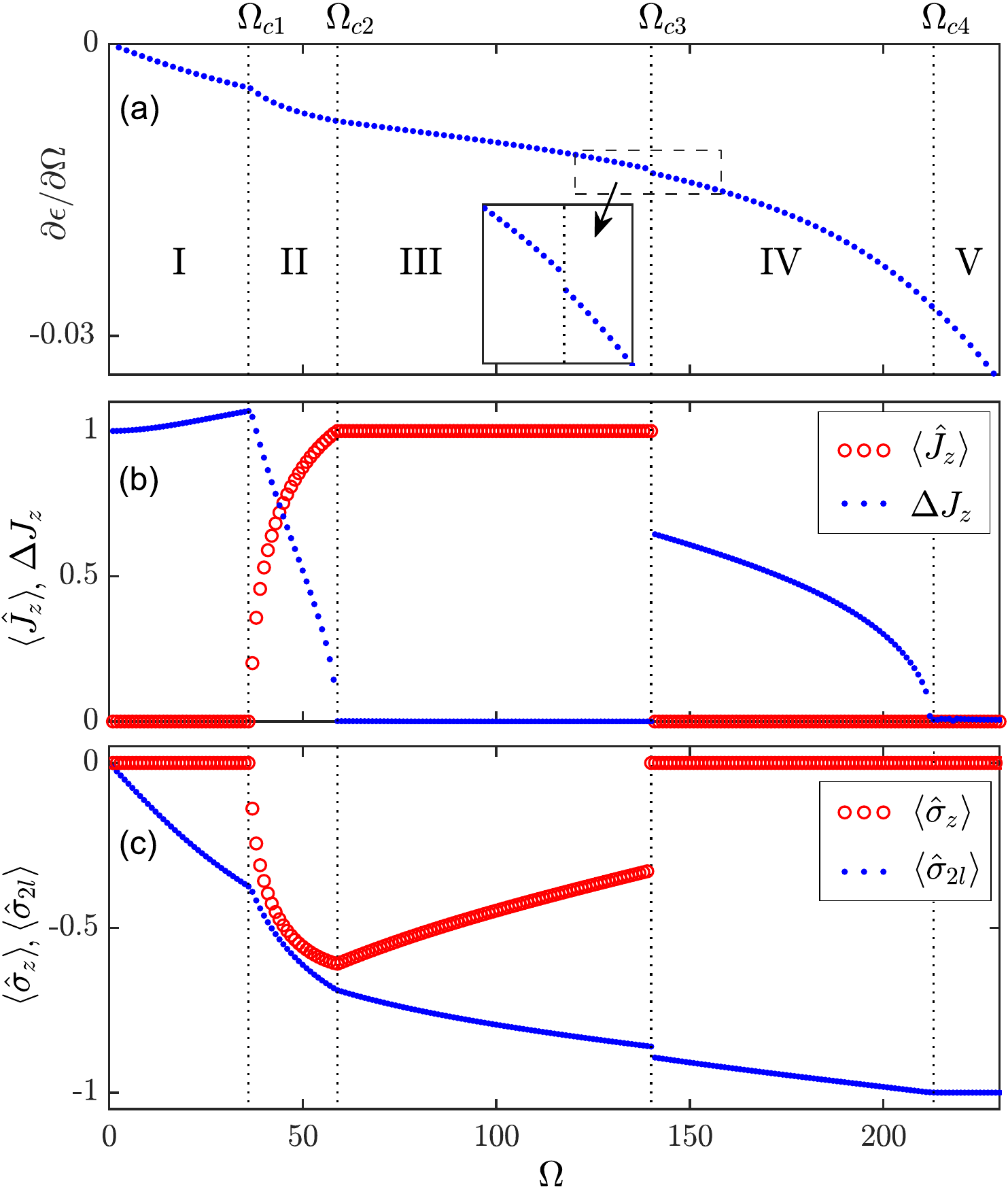}
	\caption{Phase diagram with respect to the coupling strength $\Omega$. Panel (a) shows the derivative of ground-state energy $\partial\epsilon/\partial\Omega$. Panel (b) shows the expectation of total angular momentum $\langle\hat{J}_z\rangle$ and its standard deviation $\Delta J_z$. Panel (c) shows expectation values of the spin operators $\langle\hat{\sigma}_z\rangle$ and  $\langle\hat{\sigma}_{2l}\rangle$. We show here only the non-negative branch of $\langle\hat{J}_z\rangle$ and $\langle\hat{\sigma}_z\rangle$, which corresponds to one of the two degenerate ground states.
		Phase transitions are detected at critical $\Omega$ values $\Omega_{c1}=36$, $\Omega_{c2}=59$, $\Omega_{c3}=140$ and $\Omega_{c4}=213$.}
	\label{fig:phasediagram1}
\end{figure}

The different types of ground-state wave functions displayed in Fig.~\ref{fig:5phases} can be unambiguously related to the symmetries of single-particle Hamiltonian. To this end, we show in Fig.~\ref{fig:phasediagram1} the previously introduced symmetry indicators $\langle\hat{J}_z\rangle$, $\Delta J_z$ and $\langle\hat{\sigma}_z\rangle$ depending on the coupling strength $\Omega$. We identify in total four critical values $\Omega_{c1},...,\Omega_{c4}$, at which both the behavior of these symmetry indicators and the spatial distribution of the wave function sharply change.
These changes represent quantum phase transitions and therefore, Fig.~\ref{fig:phasediagram1} can be seen as a phase diagram showing five phases (I--V) of the system, whose corresponding wave functions are already shown in Fig.~\ref{fig:5phases}.

The phase transitions identified in Fig.~\ref{fig:phasediagram1} can be classified by their orders.
Adopting the Ehrenfest classification, the order of a transition is the order of the lowest derivative of energy which shows a discontinuity \cite{landau2013statistical}.
The first-order and second-order (continuous) phase transitions significantly differ in a way how the ground state changes with the variation of a control parameter.
For the first-order transition, the ground state switches from the stationary state of one phase to that of the other phase due to an energy crossing between two states. In contrast, the continuous phase transition is characterized by the smooth transformation of the ground state. For this case, the ground-state wave functions of two phases are indistinguishable at the transition point. 
In the present study, to determine the order of transitions, we calculate the derivative of energy with respect to the coupling strength, $\partial\epsilon/\partial\Omega$. As seen in Fig.~\ref{fig:phasediagram1}(a), this quantity shows that only the transition between phases (III) and (IV) is of the first order, while all other phase transitions are continuous.
This observation is quite surprising,
as previous works have shown that transitions in SOAM coupled BECs are mostly of the first-order \cite{PhysRevA.91.053630,PhysRevA.91.063627,PhysRevLett.122.110402,PhysRevResearch.2.033152}.
Below we will argue that this behavior is explained by the existence of phases (II) and (IV), which possess neither of the symmetries of the Hamiltonian (\ref{eq:ham}) and exhibit a nontrivial vortex structure. In particular, their wave functions contain the so-called ``vortex molecule", i.e., two vortices located at a distance from each other. In order to better understand the role of such molecule in the observed continuous transitions, we first need to discuss the properties of individual phases: (I), (III) and (V) in Sec.~\ref{sub:symmetricphases} and (II) and (IV) in Sec.~\ref{sec:vortex-molecule}. In this discussion of various phases, special attention will be paid to the vortex structure of wave function and its relation to the symmetry indicators.
Moreover, for the sake of bookkeeping, we summarize the properties of all five phases in Table \ref{tab:table1}. 

\begin{table*}
\caption{\label{tab:table1} Properties of the ground state in phases (I--V). 
The upper part shows the preserved symmetries and values of three symmetry indicators. 
T and R denote the time-reversal and rotational symmetries, respectively. The lower part shows features of the wave function. In the last row, the total phase winding numbers of up and down components are shown in brackets. 
}
\begin{ruledtabular}
\begin{tabular}{cccccc}
    &I&II&III&IV&V\\[-0.1cm]
    &\begin{tabular}{@{}c@{}}Stripe\\
    phase\end{tabular}&\begin{tabular}{@{}c@{}}Two-vortex-\\
    molecule phase\end{tabular}&\begin{tabular}{@{}c@{}}Polarized\\
    phase\end{tabular}&\begin{tabular}{@{}c@{}}Vortex-antivortex-\\
    molecule phase\end{tabular}&\begin{tabular}{@{}c@{}}Zero-momentum\\
    phase\end{tabular}\\\hline
    Symmetry&T&$\boldsymbol{-}$&R&$\boldsymbol{-}$&T, R\\
    $\langle \hat{\sigma}_z \rangle$&0&$\neq0$&$\neq0$&0&0\\
    $\Delta J_z$&$\gtrsim 1$&$\neq0$&0&$\neq0$&0\\
    $\langle \hat{J}_z \rangle$&$0$&non-integer&$1$ or $-1$&0&0\\\hline
    Degeneracy&$\boldsymbol{-}$&two-fold&two-fold&$\boldsymbol{-}$&$\boldsymbol{-}$\\
    \begin{tabular}{@{}c@{}}Corresponding single-\\
    particle states\end{tabular}&$\Phi_{1-}$\,[see Eq.~(\ref{eq:superosition})]&$\boldsymbol{-}$&$\Psi_1$ or $\Psi_{-1}$ [see Eq.~(\ref{eq:jeigf})]&$\boldsymbol{-}$&$\Psi_0$\,[see Eq.~(\ref{eq:jeigf})]\\
    Angular density stripes&$\boldsymbol{+}$&$\boldsymbol{+}$&$\boldsymbol{-}$&$\boldsymbol{+}$&$\boldsymbol{-}$\\
    Phase winding number&$(0,0)$&$(+2,0)$ or $(0,-2)$&$(+2,0)$ or $(0,-2)$&$(+1,-1)$&$(+1,-1)$\\
\end{tabular}
\end{ruledtabular}
\end{table*}

\subsection{Symmetric phases}\label{sub:symmetricphases}

We refer to phases (I), (III) and (V) as the symmetric phases, since each of them possesses either one or both of the symmetries of the system. These symmetries can be clearly identified from the symmetry indicators in Fig.~\ref{fig:phasediagram1}. Specifically, vanishing spin polarization $\langle\hat{\sigma}_z\rangle$ in phases (I) and (V) implies the time-reversal symmetry of the ground state, while rotational symmetry can be identified in phases (III) and (V) by $\Delta J_z=0$. 

Based on the symmetry properties, it is straightforward to show that phases (I), (III) and (V) are the well-known stripe, polarized and zero-momentum phases that have been thoroughly studied in systems with SOAM or SLM coupling \cite{PhysRevLett.108.225301,PhysRevLett.105.160403,PhysRevA.91.053630,PhysRevA.91.063627}. 
Moreover, a correspondence can be established between each of these phases and the single-particle state with the same symmetry, as seen in Table \ref{tab:table1}.
For phases (III) and (V), their wave functions resemble Eq.~(\ref{eq:jeigf}). However, for the stripe phase, $\Phi_{1-}$ defined by Eq.~(\ref{eq:superosition}) is only a rough approximation of the many-body ground state, as suggested by $\Delta J_z\gtrsim 1$.

Before preceding to the discussion of other phases, we would like to comment on the completeness of the phase diagram. It leads us to the question whether any phases can exist beyond phase (V). 
To answer this question, we consider the strong-coupling limit ($\Omega\to\infty$) of the system, in which the total Hamiltonian $\hat{H}$ is dominated by the Raman coupling term, $\hat{H}\approx\tilde{\Omega}(\rho)\hat{\sigma}_{2l}$.
Consequently, the ground state in this regime is an eigenstate of the unitary operator $\hat{\sigma}_{2l}$ corresponding to its lowest eigenvalue $-1$.
As seen from Fig.~\ref{fig:phasediagram1}, this condition is already fulfilled in phase (V), for which $\langle\hat{\sigma}_{2l}\rangle=-1$. 
We conclude, therefore, that phase (V) represents the strong-coupling limit of the system, and no other phases
are expected for $\Omega>\Omega_{c4}$. 

\subsection{Vortex-molecule phases}\label{sec:vortex-molecule}

The phase diagram in Fig.~\ref{fig:phasediagram1} shows that, the three symmetric phases are separated by phases (II) and (IV). These two intermediate phases have no counterparts in the single-particle spectrum and also do not exist in systems with SLM coupling. 
In this section, we will characterize these two phases and reveal the role of vortex molecule in continuous phase transitions. 

\subsubsection{Two-vortex-molecule phase}\label{vortexvortexpair}

The phase (II) is observed in a relatively narrow region between the $\mathrm{T}$-symmetric stripe phase and $\mathrm{R}$-symmetric polarized phase. The ground-state wave function of phase (II) is, however, neither $\mathrm{T}$- nor $\mathrm{R}$-symmetric, as suggested by nonzero values of $\langle\hat{\sigma}_z\rangle$ and $\Delta J_z$.
Therefore, two continuous phase transitions (I--II) and (II--III) are associated with broken time-reversal and rotational symmetry, respectively.
Following a common practice for describing such transitions, we define the so-called order parameter, which is zero in symmetric phase and non-zero in symmetry-broken phase.
In our study, the symmetry indicators can be used as such order parameters. For instance, $\Delta J_z$, which continuously changes from nonzero in phase (II) to zero in phase (III), is an order parameter for the transition (II--III). Similarly, transition (I--II) is characterized by $\langle\hat\sigma_z\rangle$ or $\langle\hat{J}_z\rangle$.

\begin{figure}[tbp]
    \centering
	\includegraphics[width=\linewidth]{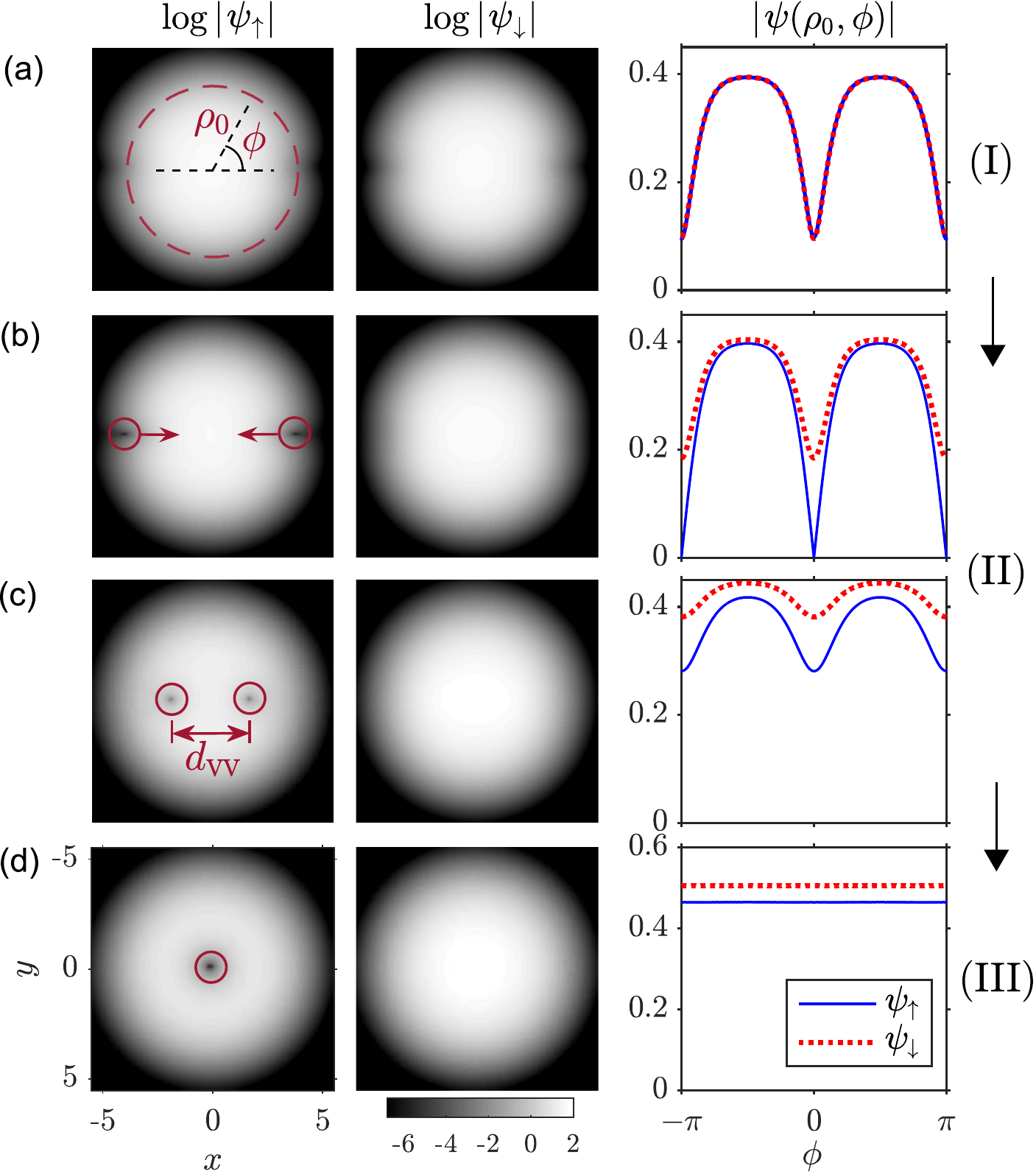}
	\caption{Amplitudes of two wave-function components (two left columns) and corresponding azimuthal distributions (rightmost column) at the distance $\rho_0$ from the center (marked with a dashed circle). Four rows correspond to different values of the coupling strength: $\Omega=36$ (a), $\Omega=37$ (b), $\Omega=45$ (c), $\Omega=61$ (d). Row (a) corresponds to the phase (I), (b) and (c) to the phase (II), (d) to the phase (III). Vorex cores are marked with circles. The value of $\rho_0$ is chosen to pass through the vortex cores in the row (b). Characteristic size of two-vortex molecule is labelled as $d_\mathrm{VV}$. The logarithmic scale is applied in two left columns for a better visibility. 
	}
	\label{fig:pentration}
\end{figure}

In addition to the symmetry indicators, the continuous phase transitions (I$\rightarrow$II$\rightarrow$III) can be identified by the transformation of the ground-state wave functions. For demonstration, we show in Fig.~\ref{fig:pentration} the wave-function amplitudes $|\psi_{\uparrow}|$ and $|\psi_{\downarrow}|$ calculated at different $\Omega$ across the two transitions. Moreover, we include in the figure the angular dependencies of these amplitudes for an arbitrarily chosen radial position $\rho_0$.
In order to describe the transitions, we start with Fig.~\ref{fig:pentration}(a), which displays the wave function of phase (I). As expected for this T-symmetric stripe phase, $|\psi_\uparrow|=|\psi_\downarrow|$ and both components exhibit azimuthal density modulations, usually referred to as stripes. The transition (I--II) is marked by the formation of vortices inside these stripes in one component and simultaneous healing of the stripes of the other, as seen in Fig.~\ref{fig:pentration}(b). This behavior leads to the fact that $|\psi_\uparrow|\neq|\psi_\downarrow|$, which shows a breaking of T-symmetry. 
As $\Omega$ increases further, the two vortices approach each other, and finally merge when the system experiences transition (II--III). At the same time, the azimuthal density modulations in both components disappear and R-symmetry is recovered, as expected for the phase (III).

\begin{figure}[tbp]
    \centering
	\includegraphics[width=0.85\linewidth,clip]{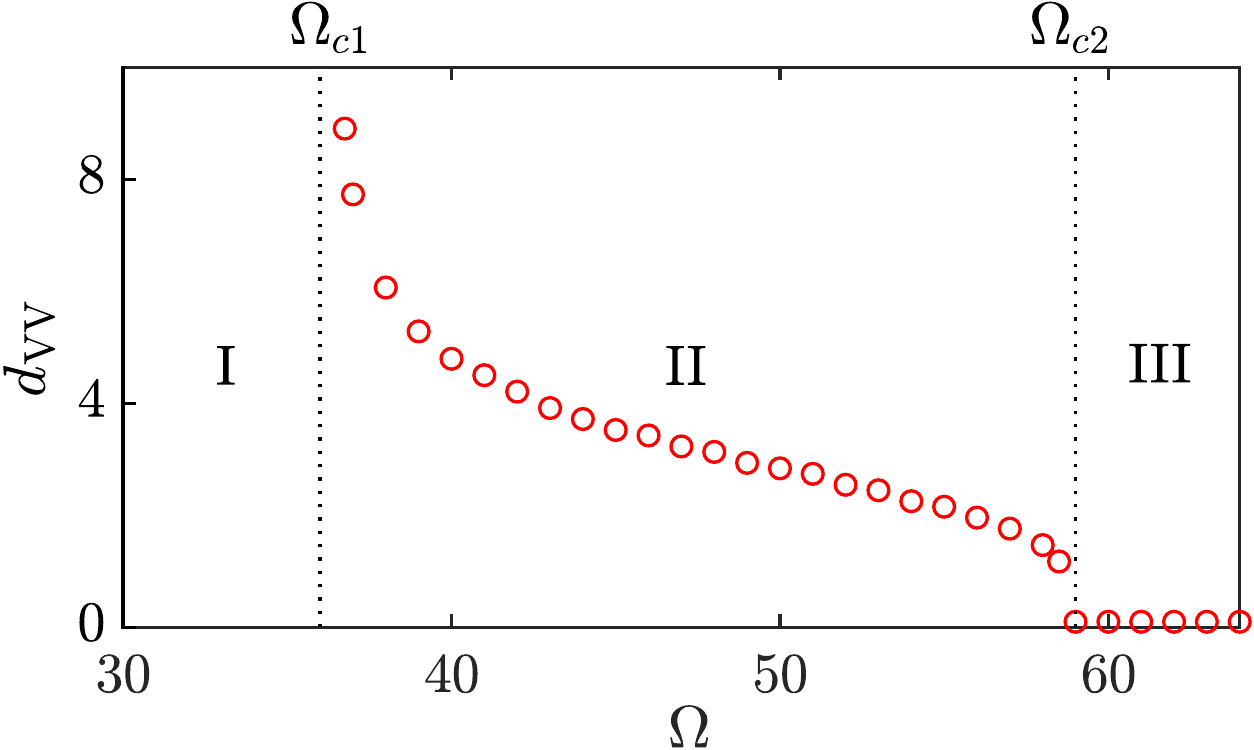}
	\caption{Vortex molecule size $d_\mathrm{VV}$ as a function of the coupling strength $\Omega$. 
	}
	\label{fig:vortexpairsize}
\end{figure}

In order to gain more insight into the role of vortices in continuous phase transitions, we show in Fig.~\ref{fig:vortexpairsize} the characteristic size of the vortex molecule, $d_\mathrm{VV}$, as a function of the coupling strength $\Omega$. Defined as the distance between two vortices, $d_\mathrm{VV}$ diverges at $\Omega_{c1}$, monotonically decreases throughout the phase (II) and becomes zero in the phase (III).
We argue therefore, that $d_\mathrm{VV}$ can serve as an alternative order parameter for the phase transition (II--III) besides the symmetry indicator $\Delta J_z$. In contrast to the latter, $d_\mathrm{VV}$ can be more accessible for experimental measurements.

\subsubsection{Vortex-antivortex-molecule phase}\label{vortexantivortexpair}

\begin{figure}[tbp]
	\centering
	\includegraphics[width=0.85\linewidth,clip]{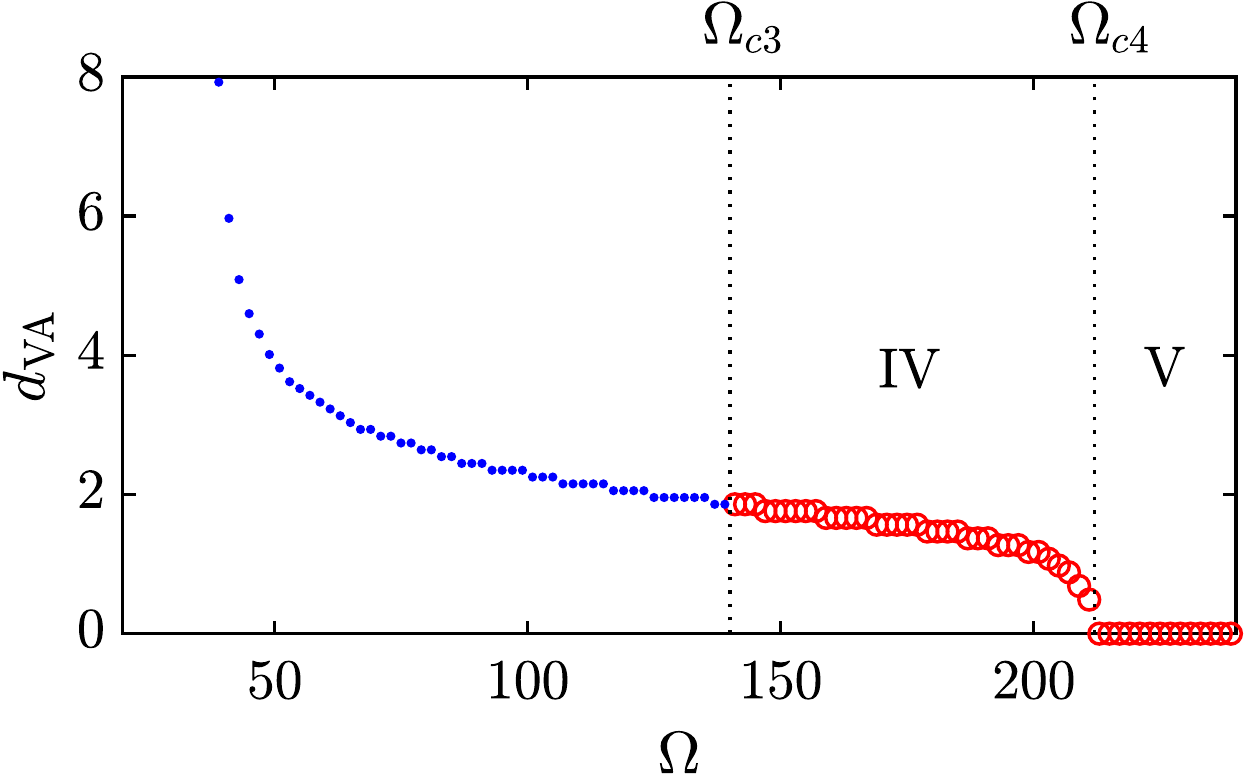}
	\caption{Vortex-antivortex molecule size $d_\mathrm{VA}$ as a function of Raman coupling strength $\Omega$. Blue dots represent the (excited) stationary vortex-antivortex-molecule state in the region $\Omega<\Omega_{c3}$.}
	\label{fig:meronpairsize}
\end{figure}

Apart from the two-vortex molecule in phase (II), a different vortex configuration can be detected in the ground state of phase (IV). 
In the latter case, each condensate component contains a single vortex. These vortices possess opposite phase windings and are displaced from each other, thus forming a vortex-antivortex molecule [see Fig.~\ref{fig:5phases}(IV)]. 
The size of this molecule, $d_\mathrm{VA}$, is displayed as a function of the coupling strength $\Omega$ in Fig.~\ref{fig:meronpairsize}. Similarly to transition (II--III), we see the collapse of vortex-antivortex molecule, $d_\mathrm{VA}=0$, at the formation of R-symmetric phase (V). However, the behavior of $d_\mathrm{VA}$ at the other phase boundary $\Omega_{c3}$ drastically differs from the divergence of $d_\mathrm{VV}$ at $\Omega=\Omega_{c1}$ (see Fig.~\ref{fig:vortexpairsize}). This difference appears, because the transition (III--IV) is a first-order transition characterized by a sudden switch of the ground state between stationary states of phases (III) and (IV). Therefore, while it is possible to trace the value of $d_\mathrm{VA}$ to $\Omega<\Omega_{c3}$, it does not describe any more the ground state of the system. 

Similarly to Sec.~\ref{sec:vortex-molecule}, the characteristic size of the vortex-antivortex molecule, $d_\mathrm{VA}$, can be used as an order parameter for the transition (IV--V). Likewise, an alternative choice can be the symmetry indicator $\Delta J_z$. However, in contrast to the phase (II), $\langle\hat{\sigma}_z\rangle$ fails to indicate the broken T-symmetry in the phase (IV), and, thus, can not serve as an order parameter. The possibility of such behavior was already mentioned in Sec.~\ref{sub:symmetry}. 

\section{summary and outlook}\label{summary}
In conclusion, we have provided a systematic description of the quantum phases and phase transitions in a two-component BEC with spin-orbital angular momentum coupling.
By analyzing the ground-state wave function for different values of the Raman coupling strength, we identified in total five quantum phases. Three of them possess either one or both of the time-reversal and rotational symmetries and represent the stripe, polarized and zero-momentum phases that are well-known in systems with spin-linear-momentum coupling. In contrast, the other two phases have no counterparts and possess neither of the above symmetries. Moreover, complex vortex-molecule structures were observed in the wave functions of these two symmetry-broken phases. We argue that the formation and collapse of vortex molecules play an important role in the phase transitions of the system. In particular, the existence of these molecules can explain the presence of continuous phase transitions in SOAM coupled condensates.
Since these transitions were not in the focus of previous studies, we carried out detailed calculations for investigating their properties. 
Our results showed that the symmetry indicators, as well as the characteristic size of the vortex molecules, can serve as effective order parameters characterizing these continuous phase transitions. 

The experimental investigation of the predicted vortex-molecule phases is feasible with currently existing techniques.
For example, the finite spin polarization, which is a indication of broken time-reversal symmetry, exhibits itself in dipole oscillations of the condensate \cite{PhysRevLett.109.115301}. Moreover, the vortex structure can be detected by interferometric measurements \cite{PhysRevLett.122.110402,PhysRevLett.102.030405} or potentially with the Bragg-scattering technique, which has been already applied for probing the stripe phase under SLM coupling \cite{Li2017,PhysRevLett.124.053605}.

The present study is restricted to the Raman coupling generated by LG beams with a phase winding $l=1$. However, the developed theory can be readily extended to higher values of $l$. Following our analysis, no additional symmetries appear in these cases, but more complicated multi-vortex molecules may exist in the condensate (see e.g. \cite{PhysRevA.91.053630}). 
A systematic investigation of such multi-vortex molecules and associated phase transitions will be presented in future publications.

\begin{acknowledgments}
Y.D. gratefully acknowledge support by the Braunschweig International Graduate School of Metrology B-IGSM and the DFG Research Training Group GrK 1952/1 ``Metrology for Complex Nanosystems." This research was also funded by the Deutsche Forschungsgemeinschaft  (DFG, German Research Foundation)  under  Germany’s  Excellence  Strategy --  EXC--2123  QuantumFrontiers--390837967.
\end{acknowledgments}

\bibliography{SOAMC}

\end{document}